\documentclass[12pt]{iopart}
\usepackage{graphicx,amsfonts}
\def\la{\left\langle}
\def\ra{\right\rangle}
\begin{document}
\title{Dynamic Multiscaling in Turbulence}
\author{Rahul Pandit$^{1,2}$\footnote{Author to whom all 
correspondence should be addressed}, 
Samriddhi Sankar Ray$^1$, and Dhrubaditya Mitra$^{3}$}
\address{$^1$ Centre for Condensed Matter Theory, Department of
Physics, Indian Institute of Science, Bangalore 560012, India.\\
$^2$ Also at, Jawaharlal Nehru Centre for Advanced Scientific Research, Bangalore 560064, India.\\
$^3$ Astronomy Unit, School of Mathematical Sciences, Queen's Mary College, London, United Kingdom. }
\ead{rahul@physics.iisc.ernet.in} 
\begin{abstract}
We give an overview of the progress that has been made
in recent years in understanding the dynamic multiscaling of
homogeneous, isotropic turbulence and related problems. We
emphasise the similarity of this problem with the dynamic scaling
of time-dependent correlation functions in the vicinity of a 
critical point in, e.g., a spin system. The universality of
dynamic-multiscaling exponents in fluid turbulence is 
explored by detailed simulations of the GOY shell model for
fluid turbulence.
\end{abstract}
\pacs{47.27.Gs, 47.53.Eq }
\section{Introduction}
\label{intro}
The statistical properties of fully developed, homogeneous, 
isotropic turbulence are often characterised by velocity
structure functions, which are averages of the differences of
fluid velocities at two points separated by a distance $r$ (a  
precise definition is given below). If $r$ lies in the 
{\it inertial range} of scales that lie between the large length 
scale $L$, at which energy is pumped into a turbulent fluid, and the 
small dissipation scale $\eta_d$, at which viscous dissipation 
becomes significant, these structure functions scale as a power of 
$r$, in a manner that is reminiscent of the algebraic behaviour of 
correlation functions at a critical point in, say, a spin system. 
This similarity between the statistical properties of homogeneous, 
isotropic turbulence and a system at a critical point is well 
known; and its elucidation for {\it equal-time} structure functions 
has been the subject of many papers: It turns out that the simple 
scaling we are accustomed to at most critical points must be 
generalised to multiscaling in turbulence; i.e., an infinity of 
exponents is required to characterise the inertial-range behaviours
of structure functions of different orders~\cite{frisch}.

The scaling behaviour of correlation functions at a critical 
point~\cite{chaikin} is associated with the divergence of a 
correlation length $\xi$ at the critical point; e.g., in a spin 
system $\xi \sim \bar{t}^{\nu}$ if the field $H = 0$, where the 
reduced temperature $\bar t\equiv (T - T_c)/T_c$, $T$ is the 
temperature, $T_c$ the critical temperature,  and $\nu$ is an 
equal-time critical exponent that is universal (for systems in a 
given universality class). In the vicinity of a critical point the 
relaxation time $\tau$, which can be determined from 
{\it time-dependent} correlation functions, scales as follows: 
\begin{equation} 
\tau \sim \xi^{z}. 
\label{ds}
\end{equation}
This is known as the {\it dynamic-scaling Ansatz} via which we 
define $z$, the dynamic-scaling exponent. Over the past few
years there has been considerable progress 
~\cite{lvov3,mitra1,mitra2,belinicher,kaneda,hayot1,hayot2} in 
developing the analogue of such a dynamic-scaling {\it Ansatz} for 
{\it time-dependent} structure functions in homogeneous,
isotropic turbulence. We give an overview of this work on 
{\it dynamic mutiscaling} in turbulence. We show, in particular,
that (a) an infinity of dynamic-multiscaling exponents is 
required here, (b) these exponents depend on the precise 
way in which relaxation times are extracted from time-dependent
structure functions, and (c) that dynamic-multiscaling exponents
are related by {\it linear bridge relations} to equal-time
multiscaling exponents.

The remaining part of this paper is organised as follows: Section 2 
contains an introduction to the multiscaling of equal-time structure
functions in fluid turbulence. Section 3 is devoted to the dynamic 
multiscaling of time-dependent structure functions. In Sect. 4 we 
introduce the GOY shell model~\cite{frisch,gledzer,ohkitani} for fluid 
turbulence and give the details of our numerical studies of this model. 
Section 5 gives representative results from our simulations 
for time-dependent structure functions for the GOY shell model
with conventional viscosity; we also present new results with
hyperviscosity and show explicitly that dynamic multiscaling
exponents are independent of the type of viscosity we use.
We end with concluding remarks about the possibility
of experimental verifications of our predictions and 
generalizations to other types of turbulence such as 
passive-scalar turbulence\cite{rmp}. 

\section{Equal-time Multiscaling} 
Fluid flows are described by the Navier-Stokes equation for the 
velocity field ${\bf u}({\bf x},t)$ at point ${\bf x}$ and time $t$: 
\begin{equation}
\partial_t{\bf u} + {\bf u}.\nabla{\bf u} = -\nabla P + 
\nu_0\nabla^2{\bf u} + {\bf f}; 
\label{NS1}
\end{equation}
we consider low-Mach number flows that are nearly incompressible and so 
equation (\ref{NS1}) must be augmented by the incompressibility 
constraint
\begin{equation}
\nabla .{\bf u} = 0,
\label{NS2}
\end{equation}
which can be used to eliminate the pressure $P$ in equation 
(\ref{NS1}); we choose the uniform density $\rho=1$; $\nu_0$ is the 
kinematic viscosity; and for decaying turbulence the external 
force ${\bf f}$ is zero. Turbulence occurs when the Reynolds number 
$Re \equiv (\ell v)/\nu_0$ is large; here $\ell$ and $v$ are 
characteristic length and velocity scales of the flow.

The chaotic nature of turbulence has led to a natural statistical 
description of the velocity field in terms of velocity 
structure functions. For instance, we can define the order-$p$, 
equal-time, structure function for longitudinal velocity 
increments $\delta u_{\parallel}({\bf x},{\bf r},t) 
\equiv [{\bf u}({\bf x}+{\bf r},t) - {\bf u}({\bf x},t)].{\bf r}/r$ as 
follows:
\begin{eqnarray}
{\cal S}_p(r) \equiv \la [\delta u_{\parallel}({\bf x},{\bf r},t)]^p \ra \sim r^{\zeta_p};
\label{Sp}
\end{eqnarray}
the power-law dependence on $r$, which defines the exponent 
$\zeta_p$, holds for $L \ll r \ll \eta_d$; and the angular brackets 
indicate an average over either the statistical steady state, 
for forced turbulence, or statistically independent initial 
conditions, for decaying turbulence.

In contrast with the scaling behaviour of correlation functions
at conventional critical points in equilibrium statistical 
mechanics, in turbulence the structure functions ${\cal S}_p(r)$ do
not exhibit simple scaling forms: Experimental and numerical 
evidence suggests that ${\cal S}_p(r)$ show multiscaling, 
with $\zeta_p$ a nonlinear, convex, monotone increasing function
of $p$~\cite{frisch}. The 1941 theory of Kolmogorov (K41)
~\cite{K41a,K41b} yields simple scaling with $\zeta_p^{K41} = p/3$; 
the measured values of $\zeta_p$ deviate significantly from 
$\zeta_p^{K41}$ for $p > 3$; and for $p=3$ we have the exact von 
K\'arm\'an-Howarth~\cite{frisch} result $\zeta_3 = 1$. 

Even though the K41 phenomenology fails to capture the 
multiscaling of the equal-time structure function ${\cal S}_p(r)$, 
it provides us with important
conceptual underpinnings for studies of homogeneous, isotropic  
turbulence. In particular, the exponents $\zeta_p^{K41}$ are 
{\it universal} in the sense that they do not depend on the
details of the dissipation, i.e., the viscosity. (Of course
measurements should be made far away from boundaries so that
the flow satisfies the conditions of homogeneity and isotropy.)
This suggestion of power-law scaling with universal scaling 
exponents was made for turbulence a few decades before it was 
appreciated fully in the context of critical phenomena.

The universality of the exponents $\zeta_p$ holds even if we go
beyond K41 phenomenology. Let us first examine the dependence
of these exponents on the dissipation mechanism. In numerical 
simulations it is possible to introduce a hyperviscosity 
$\nu_{\alpha}$ by replacing the viscous term $\nu_0\nabla^2{\bf u}$ 
in equation (\ref{NS1}) by $\nu_{\alpha}\nabla^{2 + \alpha}{\bf u}$;
here $\alpha \geq 0$ determines the {\it degree} of hyperviscosity; 
and $\alpha = 0$ yields normal viscous dissipation. 
Some early shell-model studies~\cite{she,schorg,ditlev}
suggested that the exponents $\zeta_p$ depend on $\alpha$.
But subsequent direct numerical simulations (DNS)~\cite{borue} 
of equations (\ref{NS1}-\ref{NS2}) and numerical studies of a shell 
model~\cite{lvov1} for turbulence have argued against this. Our 
results are consistent with these later studies.

The exponents $\zeta_p$ do not seem to depend on whether they are 
measured in statistically steady, forced turbulence or in 
decaying turbulence. Numerical evidence for this 
{\it universality} has been provided by the shell-model 
studies of Ref.~\cite{lvov2}.

In the remaining Sections of this paper we will show how to 
generalise the dynamic-scaling {\it Ansatz} (\ref{ds}) to account for
multiscaling in turbulence. Our discussion will be based on 
earlier studies~\cite{lvov3,belinicher} and the work carried out in our 
group~\cite{mitra1,mitra2}. We will then examine the universality of 
the dynamic-multiscaling exponents in fluid turbulence, i.e.,
we will explore their dependence on (a) the hyperviscosity 
parameter $\alpha$ and (b) the type of turbulence (statistically
steady as opposed to decaying). 

\section{Dynamic Multiscaling}

The dynamic scaling of time-dependent correlation functions at a 
critical point in an equilibrium system was systematised soon after 
the scaling of equal-time correlation functions. The analogous
development for homogeneous, isotropic turbulence has been 
carried out recently~\cite{lvov3,mitra1,mitra2,belinicher,kaneda}.
We summarise the essential points here before presenting our new
results. A na\"ive extension of K41 phenomenology to time-dependent 
structure functions yields a dynamic exponent $z^{K41} = 2/3$ for 
all orders $p$. Two improvements are required to go beyond this 
K41 result: (a) We must account for the multiscaling of velocity 
structure functions. (b) We must distinguish between structure 
functions of Eulerian and Lagrangian velocities; the former yield 
trivial dynamic scaling with $z^{\cal E} = 1$, for all $p$, since 
the mean flow (or the flow of the largest eddy) directly advects 
small eddies and so temporal and spatial separations are related 
linearly by the mean flow velocity; nontrivial dynamic multiscaling 
can be anticipated, therefore, only for Lagrangian~\cite{pope} or 
quasi-Lagrangian~\cite{belinicher} velocity structure functions. 
The latter are defined in terms of the quasi-Lagrangian velocity 
${\bf \hat u}$ that is related to its Eulerian counterpart $\bf{u}$ as 
follows: 
\begin{equation}
{\bf \hat u}({\bf x},t) 
\equiv {\bf u}[{\bf x} + {\bf R}(t;{\bf r_0},0),t] ,
\label{qltrans}
\end{equation}
with ${\bf R}(t;{\bf r_0},0)$ the position at time $t$ of a 
Largrangian particle that was at ${\bf r_0}$ at time $t = 0$.
Equal-time, quasi-Lagrangian velocity structure functions 
are the same as their Eulerian counterparts~\cite{lebdev}.

The order-$p$, time-dependent, structure function, for longitudinal, 
quasi-Lagrangian velocity increments is~\cite{lvov3,mitra1} 
\begin{eqnarray}
{\mathcal F}_p(r,\{t_1,\ldots,t_p\}) \equiv
        \la [\delta \hat{u}_{\parallel}({\bf x},t_1,r) \ldots
              \delta \hat{u}_{\parallel}({\bf x},t_p,r)] \ra.
\label{Fp}
\end{eqnarray}
Since we are interested in scaling behaviours, we restrict $r$ to 
the inertial range, consider, for simplicity, 
$t_1=t$ and $t_2=\ldots=t_p=0$, and denote the structure function 
(\ref{Fp}) by ${\mathcal F}_p(r,t)$. Given ${\mathcal F}_p(r,t)$, 
there are
different ways of extracting time scales.  For example, we can use
the following order-$p$, degree-$M$ integral- and derivative-time 
scales that are defined, respectively, as~\cite{mitra1} 
\begin{equation}
{\cal T}^{I}_{p,M}(r) \equiv
 \biggl[ \frac{1}{{\mathcal S}_p(r)}
\int_0^{\infty}{\mathcal F}_p(r,t)t^{(M-1)} dt
\biggl]^{(1/M)}
\label{timp}
\end{equation}
and
\begin{equation}
{\cal T}^{D}_{p,M}(r) \equiv
 \biggl[ \frac{1}{{\mathcal S}_p(r)}
\frac{\partial^M {\mathcal F}_p(r,t)}{\partial t^M}\biggl|_{t=0}
\biggl]^{(-1/M)}.
\label{tdp}
\end{equation}
If the integral in (\ref{timp}) and the derivative in (\ref{tdp}) 
exist we can generalise the dynamic-scaling {\it Ansatz} (\ref{ds}) 
at a critical point to the following dynamic-multiscaling
{\it Ans\"atze} for homogeneous, isotropic 
turbulence~\cite{lvov3,mitra1}:
\begin{equation}
{\cal T}^{I}_{p,M}(r) \sim  r^{z^{I}_{p,M}};
\label{zi} 
\end{equation}
\begin{equation}
{\cal T}^D_{p,M}(r) \sim  r^{z^D_{p,M}}.
\label{zd} 
\end{equation}
These equations define, respectively, the integral- and the  
derivative-time multiscaling exponents $z^{I}_{p,M}$ and 
$z^{D}_{p,M}$, which satisfy the bridge relations
\begin{equation}
z^I_{p,M} = 1 + [\zeta_{p-M} - \zeta_p]/M, 
\label{zipm}
\end{equation}
obtained first in Ref.~\cite{lvov3},
and
\begin{equation}
z^{D}_{p,M} = 1 + [\zeta_p - \zeta_{p+M}]/M ,
\label{zdpm}
\end{equation}
obtained in Ref.~\cite{mitra1}.
These bridge relations follow from a 
generalisation~\cite{lvov3,mitra1} of the multifractal 
formalism~\cite{frisch} for turbulence. Here we sketch the 
arguments that lead to equation (\ref{zipm}) for statistically
steady turbulence and refer the reader to Refs.~\cite{mitra1,ray} 
for details: We begin by assuming the following multifractal form 
for the order-$p$ time-dependent structure functions: 
\begin{equation}
\frac{{\cal F}_p(r,t)}{\hat{u}_L^p} \propto \int_Id\mu(h) 
\big{(}\frac{r}{L}\big{)}^{3 + ph 
- {\cal D}^{\hat{u}}(h)}{\cal G}^{p,h}(\frac{t}{\tau_{p,h}}).
\label{Fpforced}
\end{equation}
As in the equal-time multifractal formalism~\cite{frisch},
the scaling exponents $h \in {\cal I} \equiv [h_{min},h_{max}]$. 
Corresponding to each exponent $h$ there is a set 
${\bf \Sigma}_h \subset {\mathbb R}^3$, of fractal dimension 
${\cal D}^{\hat{u}(h)}$ and with a measure  $d\mu(h)$, such that
$\delta \hat{u}_r({\bf x})/\hat{u}_L \sim (r/L)^h$ if 
${\bf x} \in {\bf \Sigma}_h$, with $\delta \hat{u}_r({\bf x}) \equiv
|{\bf \hat{u}}({\bf x+r}) - {\bf \hat{u}}({\bf x})|$.
We assume furthermore that the scaling function 
${\cal G}^{p,h}(\frac{t}{\tau_{p,h}})$ is such that 
${\cal G}^{p,h}(0) = 1$ and that the characteristic decay time 
$\tau_{p,h} \sim r/\delta \hat{u}_r({\bf x}) \sim r^{1-h}$. 
If we substitute for ${\cal F}_p(r,t)$ in equation (\ref{zi}), do 
the time integral first by a saddle-point method, we obtain the 
bridge relation (\ref{zipm}).  Similar calculations lead to 
(\ref{zdpm}).
A complete discussion of such bridge relations, as well as similar 
relations for passive-scalar turbulence, can be found in 
Refs. ~\cite{mitra1,mitra2,ray}. The last of these references 
also shows that the bridge relations are the same for decaying and
statistically steady turbulence.

Given current computational resources, it has not been possible 
to verify the bridge relations (\ref{zipm}-\ref{zdpm}) by 
computing quasi-Lagrangian velocities in a 
direct numerical simulation of the Navier-Stokes equation
(\ref{NS1}-\ref{NS2}) for an incompressible fluid.
Thus we must turn to numerical studies of simple shell
models of turbulence that we describe in the next Section.

\section{The GOY Shell Model}

We have carried out extensive numerical simulations to obtain 
equal-time and dynamic multiscaling exponents for the GOY
shell model for fluid turbulence~\cite{frisch,gledzer,ohkitani}.  
This shell model is defined 
on a logarithmically discretized Fourier space labelled by 
scalar wave vectors $k_n$ that are associated with the shells
$n$. The dynamical variables are the complex scalar shell
velocities $u_n(k_n)$, henceforth denoted by $u_n$. 
The evolution equations for the GOY model are
\begin{eqnarray}
\hspace*{-1cm}
\left[\frac{d}{dt} + \nu_0\left(\frac{k_n}{k_d}\right)^\alpha k_n^2\right]u_n =  
\imath\bigg{[}a_n u_{n+1}u_{n+2} +   
b_n u_{n-1}u_{n+1} + c_n u_{n-1}u_{n-2} \bigg{]}^{\ast}, 
\label{goy}
\end{eqnarray}
where $k_n = k_0 2^n\/$, $k_0$ = 1/16, complex conjugation is 
denoted by $\ast$, and the coefficients $a_n = k_n$, $b_n = 
-\delta k_{n-1}$, $c_n = -(1-\delta)k_{n-2}$ are chosen to  
conserve the shell-model analogues of energy and helicity
in the inviscid, unforced limit, and $1\leq n \leq N$. We use the 
standard value $\delta = 1/2$. The second term allows for 
hyperviscous dissipation with degree $\alpha$ (the case
$\alpha = 0$ corresponds to conventional viscous dissipation); 
$k_d$ is a large wavenumber whose inverse is comparable to the 
dissipation length scale; we choose $k_d = 2^{18}$ and 
$\alpha = 0$ or $\alpha = 2$.  Since we concentrate on decaying 
turbulence here, the forcing term, required to drive the system into 
a statistically steady state, is absent.  The logarithmic 
discretization of Fourier space allows us 
to reach very high Reynolds numbers in numerical simulations
of the GOY model even with $N = 22$.

We use a slaved, second-order, Adams-Bashforth scheme
~\cite{sujan,pisarenko} to 
integrate the GOY-model equations with $N=22$ shells, and, in some 
representative cases, $N=35$ (this requires a fifth-order scheme
~\cite{sahoo} ), 
with the boundary conditions $u_n = 0$ for $n \leq 0$ and
$n > N$. In our simulations we use $\delta t = 10^{-4}$ as the 
integration time step, and the viscosity $\nu = 10^{-7}$ for 
$\alpha = 0$ and $\nu = 10^{-6}$ for $\alpha = 2$. 

The GOY-model equations allow for direct interactions only between
nearest- and next-nearest-neighbour shells. By contrast, in 
the Fourier transform of the Navier-Stokes equation every Fourier 
mode of the velocity is coupled to every other Fourier mode 
directly. This direct sweeping, as it is called, leads
to the trivial dynamic scaling of Eulerian velocity structure
functions (Sect. 3). Since the GOY model does not have direct
sweeping in this sense, it is sometimes thought of as a highly
simplified quasi-Lagrangian version of the Navier-Stokes 
equation. Thus we might expect nontrivial dynamic multiscaling for 
GOY-model structure functions. The equal-time structure functions  
for this model are
\begin{equation}
S_p(k_n) \equiv \la [u_n(t)u^{\ast}_n(t)]^{p/2} \ra \sim k_n^{-\zeta_p},
\label{goy1}
\end{equation}
where the power-law dependence is obtained only if $k_n^{-1}$ lies
in the inertial range. Three cycles~\cite{kad1}
in the static solutions of the GOY model lead to rough, 
period-three oscillations in $S_p(k_n)$; thus we use the modified 
structure function~\cite{kad1} 
\begin{eqnarray}
\Sigma_p &\equiv & \la|{\Im}[u_{n+2}u_{n+1}u_n - 
(1/4)u_{n-1}u_nu_{n+1}]|^{p/3}\ra \nonumber \\ 
&\sim & k_n^{-\zeta_p},
\label{sigma}
\end{eqnarray}
in which these oscillations are effectively filtered out.
Therefore we use equations (\ref{sigma}) and (\ref{goy2}) to extract 
$\zeta_p$. Time scales are obtained from the order-$p$, 
time-dependent structure functions for the GOY model, namely,  
\begin{equation}
F_p(k_n,t_0,t) \equiv \la [u_n(t_0)u^{\ast}_n(t_0 + t)]^{p/2} \ra.
\label{goy2}
\end{equation}

In Refs.~\cite{mitra1,ray} we have used such time-dependent 
structure functions to verify the bridge relations 
(\ref{zipm}-\ref{zdpm}) for statistically steady turbulence.
Here we give a short overview of our results for decaying turbulence
with conventional viscosity $(\alpha = 0)$ and hyperviscosity
$(\alpha = 2)$. In our studies of decaying turbulence
we have used two types of initial conditions; in both of these  
all the energy is initially concentrated in the first few Fourier 
modes, i.e., at large length scales, as in typical wind-tunnel 
experiments of homogeneous, isotropic turbulence~\cite{frisch}. In
the first initial condition $u_n = k_n^{1/2} e^{i\theta_n}$, 
for $n$ = 1, 2, and $u_n = k_n^{1/2}e^{{-k_n}^2} e^{i\theta_n}$
for $3\leq n \leq N$, where $\theta_n$ is a random phase angle 
distributed uniformly between 0 and $2\pi$; we use this for 
the case $\alpha = 0$. The second initial condition we use is 
similar, namely, $u_n = e^{-k_n^2}e^{i\theta_n}$ for 
$1\leq n \leq N$. 
We have used both these initial conditions for the case 
$\alpha = 2$; our results do not depend significantly on the 
initial condition that is used; the results we report here 
for $\alpha =2$ are for the second type of initial condition.

Time is measured in terms of the initial large 
eddy-turnover time $t_L$.  For the GOY shell model 
$t_L \equiv 1/(u_{rms}k_1)$, where the root-mean-square velocity 
$u_{rms} \equiv [\la \sum_n |u_n|^2 \ra]^{1/2}$ is 
defined for the initial velocity field. The shell-model energy 
spectrum is~\cite{sujan} $E(k_n) \equiv \la |u_n|^2 \ra/k_n 
= S_2(k_n)/k_n$; since $S_2(k_n)$ shows the period-three oscillations
mentioned above, a smooth energy spectrum is obtained by using 
$\Sigma_2(k_n)/k_n$ as we show below.  The mean kinetic energy 
dissipation rate and the integral scale for the GOY model are, 
respectively, 
$\epsilon \equiv \la \sum_n \nu(k_n/k_d)^\alpha k_n^2|u_n^2| \ra$
and $L_{int} \equiv \frac{\la\sum_n(|u_n^2|/k_n^2)\ra}{\la\sum_n(|u_n^2|/k_n)\ra}$. 
As the turbulence decays, $L_{int}$ increases;
once it becomes comparable to the size of the system the total 
energy decays~\cite{borue} as $t^{-2}$. Our results for 
dynamic-multiscaling exponents are obtained for times that are
much shorter than the time over which $L_{int}$ becomes comparable
to the system size ($\sim k_1^{-1}$ for the GOY model).  
 
\section{Results and Conclusions}
 
\begin{figure}[htbp]
\begin{center}
\includegraphics[height=6cm,width=6cm]{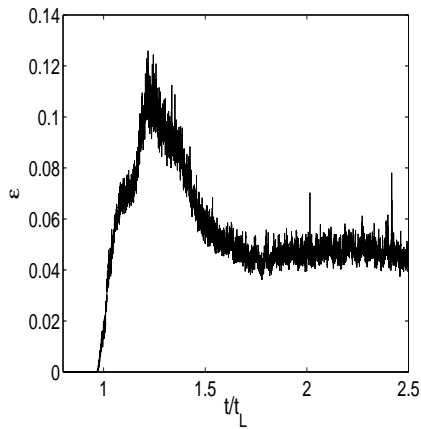}
\caption{The mean energy dissipation rate $\epsilon$ versus time for 
$\alpha = 2$ (for clarity we show data for 
$0.8 \leq t/t_L \leq 2.5$); these data have been averaged over 2000 
initial conditions. The main peak at $t/t_L \simeq 1.2$ is a 
signature of cascade completion.  For $\alpha = 0$ $\epsilon$ 
displays a similar peak.} 
\label{epsilon}
\end{center}
\end{figure}

\begin{figure}[htbp]
\begin{center}
\includegraphics[height=6cm,width=6cm]{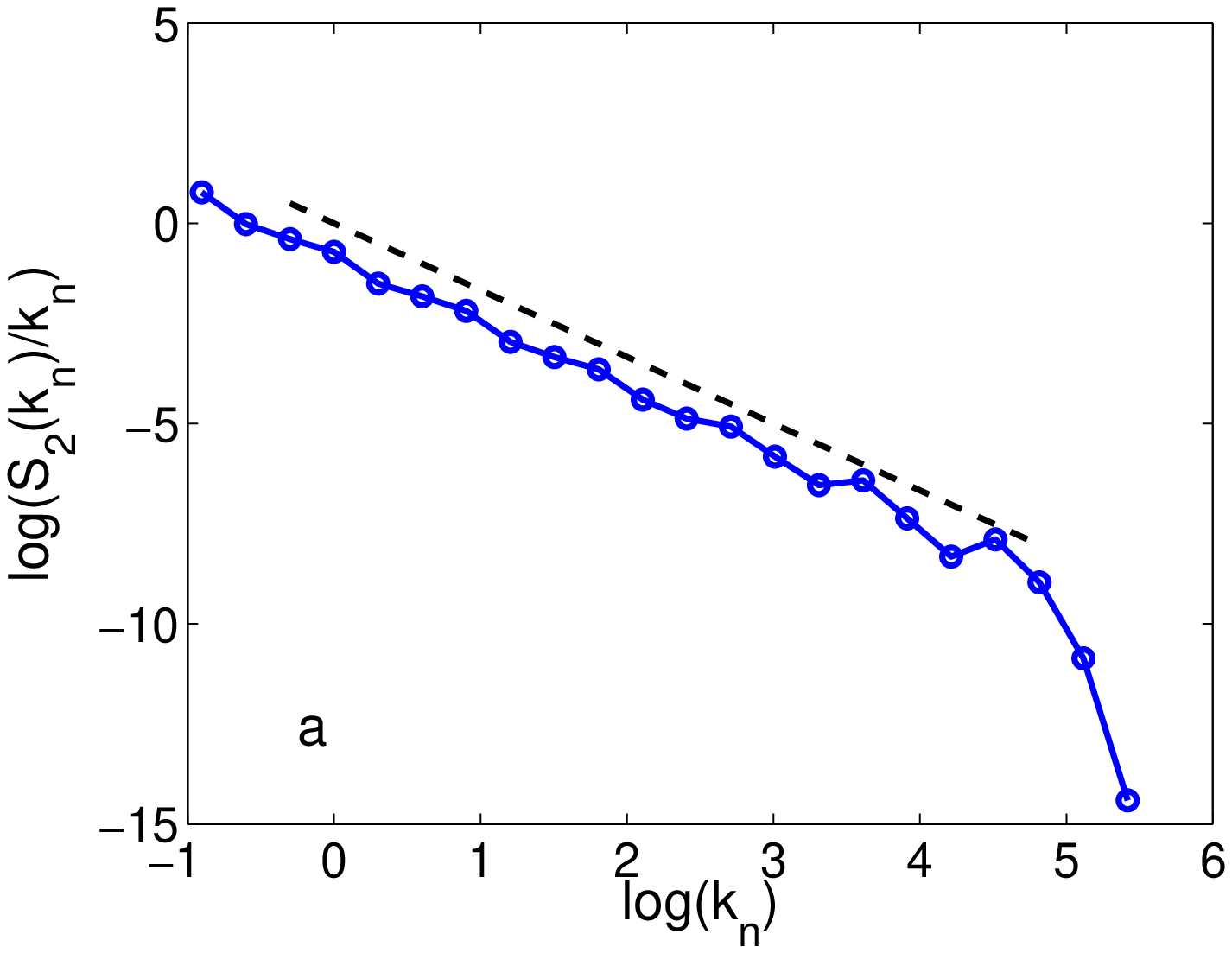}
\includegraphics[height=6cm,width=6cm]{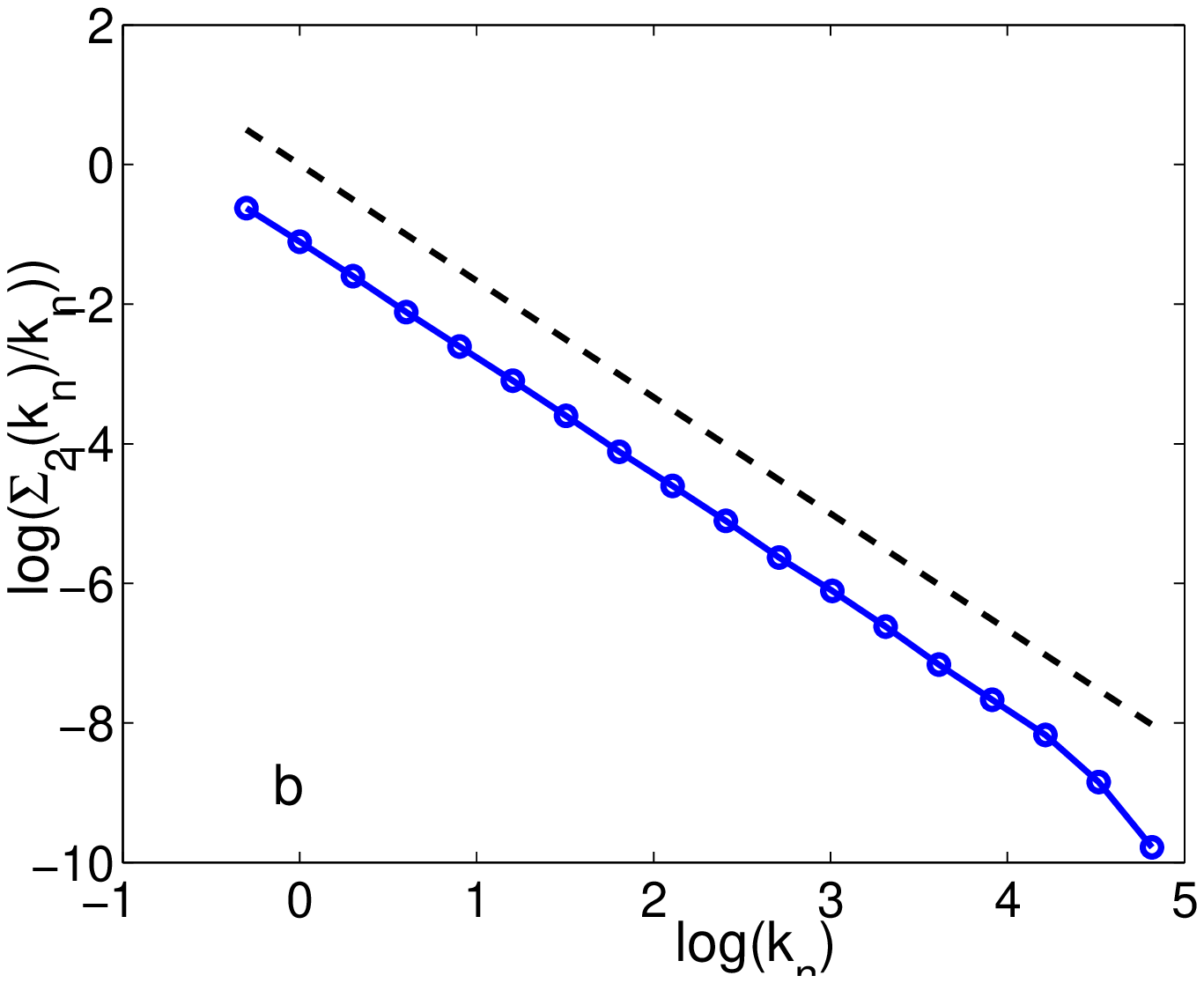}	
\caption{(a) Log-log plot of the kinetic energy spectrum  
$E(k_n) = S_2(k_n)/k_n$ versus $k_n$ with the period-three 
oscillations (see text). The black dashed line corresponds
to the K41 scaling prediction $E(k_n) \sim k^{-5/3}$. 
(b) Log-log plot of $\Sigma_2(k_n)/k_n$ versus $k_n$; note that  
the period-three oscillations are suppressed here; the black 
dashed line indicates the K41 scaling prediction.}
\label{spectrum}
\end{center}
\end{figure}

\begin{figure}[htbp]
\begin{center}
\includegraphics[height=6cm,width=6cm]{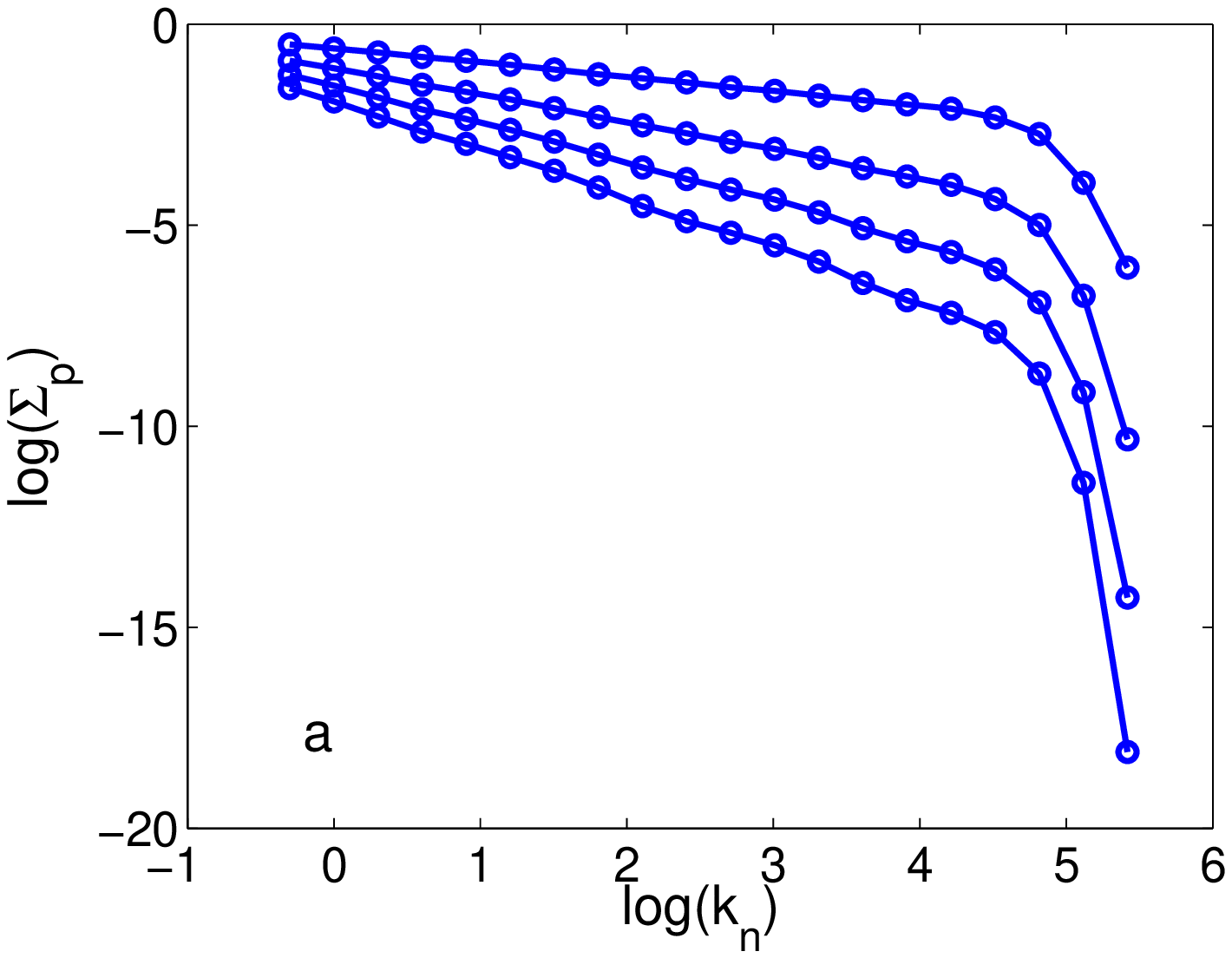}
\includegraphics[height=6cm,width=6cm]{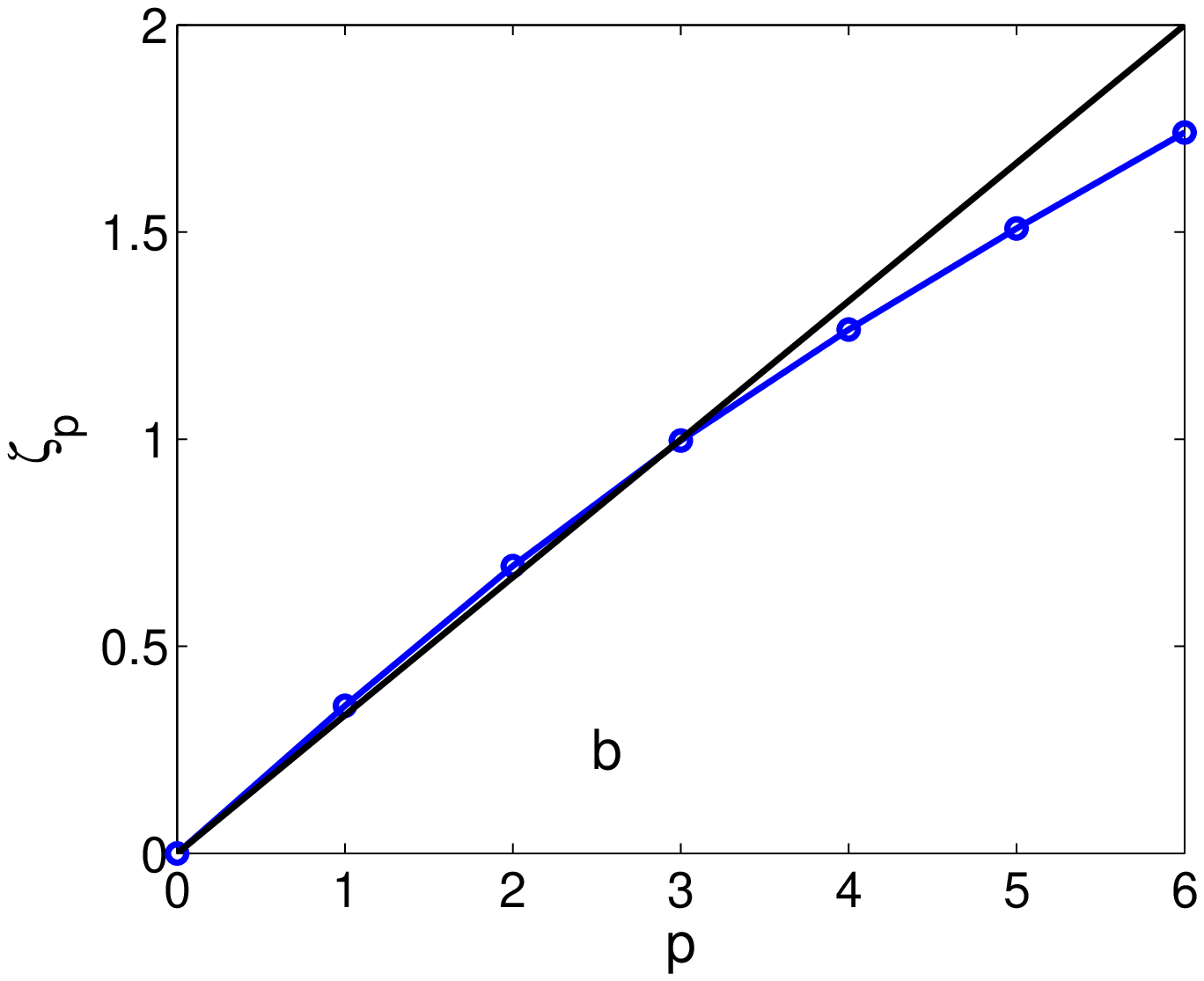}
\caption{(a) Log-log plot of the modified structure function 
$\Sigma_p$, for $p = 1$ to $4$ (top to bottom), versus $k_n$ 
for $\alpha = 2$ and number of shells $N=35$ (we show data
only for the first $22$ shells). (b) Plot of the equal-time 
multiscaling exponents $\zeta_p$ (obtained from (a) and listed in
Table 1) versus the order $p$. The open circles (o), connected by a 
curve to guide the eye, indicate data from our numerical 
simulations; the thick black line is the K41 prediction 
$\zeta_p^{K41} = p/3$.}
\label{sigmaplot}
\end{center}
\end{figure}

We begin by looking at the mean energy dissipation rate $\epsilon$ 
as a function of time. A representative plot, averaged over 2000 
initial conditions, is shown for $\alpha = 2$ in 
Figure \ref{epsilon}. Though this plot is noisy, it shows a 
clear peak. This peak, at $t/t_L \simeq 1.2$ ($t_L \simeq 5$) in 
Figure \ref{epsilon}, 
signals the completion of the cascade that transfers energy from 
the scale at which it is injected to the small scales where viscous 
dissipation becomes significant. Representative energy spectra at 
cascade completion are shown in Figure \ref{spectrum}; after this 
point in time  the energy spectrum decays very slowly without an 
{\it appreciable} change in the slope of the scaling regime, which 
appears as a nearly straight-line segment in the log-log plots
of Figure \ref{spectrum}. Straight-line segments also show up
in plots of the structure function $\Sigma_p$ at (or after) cascade
completion as shown in the representative plots, for $\alpha = 2$, 
of Figure \ref{sigmaplot}(a); each curve in this Figure has been
averaged over 5000 independent initial conditions. From the slopes 
of the straight-line segments in these plots (see Table 1)
we obtain the equal-time multiscaling exponents $\zeta_p$ whose 
dependence on $p$ is shown in Figure \ref{sigmaplot}(b). 
The values we quote for these (and other) exponents are 
the means of the slopes of 50 different plots like 
Figure \ref{sigmaplot}(a), which are obtained from 50 statistically 
independent runs; the corresponding standard deviations yield the 
error bars shown in Table 1.

Our results for the equal-time exponents $\zeta_p$ for $\alpha = 0$ 
and $\alpha = 2$ are presented in Table 1. By comparing Columns 
$2$ and $3$ in this Table we see that the exponents for both values
of $\alpha$ agree with each other and with the exponents reported
earlier for statistically steady turbulence~\cite{mitra1}. Thus we 
reconfirm the universality of 
equal-time exponents: they neither depend on the precise dissipation 
mechanism nor on whether we consider statistically steady or
decaying turbulence.  

\begin{table}
\caption{Our results for the equal-time multiscaling exponents 
$\zeta_p$ at cascade completion for $\alpha = 0$ (Column 2) and 
$\alpha = 2$ (Column 3). We indicate, in parentheses, the ranges of 
shell numbers $n$ over which we fit our data for equal-time
structure functions to obtain these exponents. Note that the 
exponents in Columns 2 and 3 agree very well with each other.}  
\label{tab:1}  
\begin{center}
\begin{tabular}{lll}
\hline\noalign{\smallskip}
order-$p$ & $\zeta_p$ (4-14)  & $\zeta_p$ (4-16) \\
\noalign{\smallskip}\hline\noalign{\smallskip}
1 & 0.380 $\pm$ 0.001 & 0.37 $\pm$ 0.01\\
2 & 0.709 $\pm$ 0.003 & 0.699 $\pm$ 0.008\\
3 & 1.000 $\pm$ 0.005 & 1.003 $\pm$ 0.008\\
4 & 1.266 $\pm$ 0.008 & 1.29 $\pm$ 0.02 \\
5 & 1.51 $\pm$ 0.01 & 1.55 $\pm$ 0.03 \\
6 & 1.74 $\pm$ 0.02 & 1.79 $\pm$ 0.05 \\	
\noalign{\smallskip}\hline
\end{tabular}
\end{center}
\end{table}

In Ref.~\cite{mitra1} the dynamic-multiscaling exponents $z^I_{p,M}$
and $z^D_{p,M}$ were obtained, by using integral- and 
derivative-time scales, for statistically steady turbulence in the 
GOY model with $\alpha=0$. We give below an overview of our recent
results for dynamic multiscaling but for decaying turbulence;
for a detailed discussion of these results we refer the reader
to Ref.~\cite{ray}.

\begin{figure}
\begin{center}
\includegraphics[height=6cm,width=6cm]{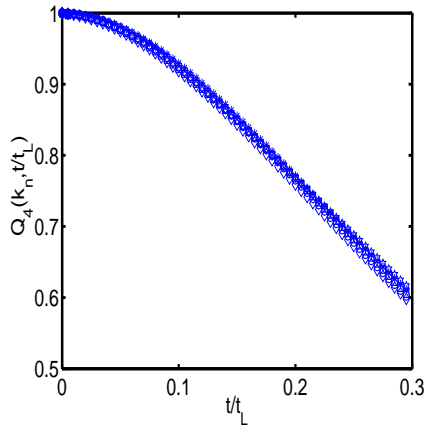}
\caption{Plots of $Q_4(k_n,t/t_L)$ for $n = 5$ obtained from  
 6 different values of $t_0$; successive values of
$t_0$ are separated by 0.5$t_L$. Since the symbols used for 
different values of $t_0$ are indistinguishable on the scale of 
this figure, we conclude that $Q_4$ does not depend on $t_0$.} 
\label{qp}
\end{center} 
\end{figure}

Time-dependent structure functions in decaying turbulence must, of 
course, depend on the origin of time $t_0$ from which we start 
making measurements. It turns out that  this dependence can
be eliminated by normalising these structure functions by
their values at the origin of time; we have shown this analytically
for the Kraichnan model~\cite{kraich1,falko} for passive-scalar 
turbulence and numerically in several other models~\cite{ray}. 
We present some representative numerical results for the GOY model 
below for which we use the normalised time-dependent structure 
functions   
\begin{equation}
Q_p(k_n,t) = \frac{F_p(k_n,t_0,t)}{F_p(k_n,t_0,0)}.
\label{Q}
\end{equation}
It turns out that $Q_p$ does not depend on $t_0$ as shown explicitly
in Figure \ref{qp} for $p=4$ and $n=5$. This is why we have
not displayed $t_0$ as an argument of $Q_p$.

Given that $Q_p$ does not depend on $t_0$, we can use 
expressions such as (\ref{zi}-\ref{zd}), with ${\cal F}_p$ 
replaced by $Q_p$, to obtain dynamic-multiscaling exponents
for decaying turbulence. We consider here the order-1 integral- 
and order-2 derivative-time scales and the associated exponents 
$z^I_{p,1}$ and $z^D_{p,2}$, respectively. 

\begin{figure}
\begin{center}
\includegraphics[height=6cm,width=6cm]{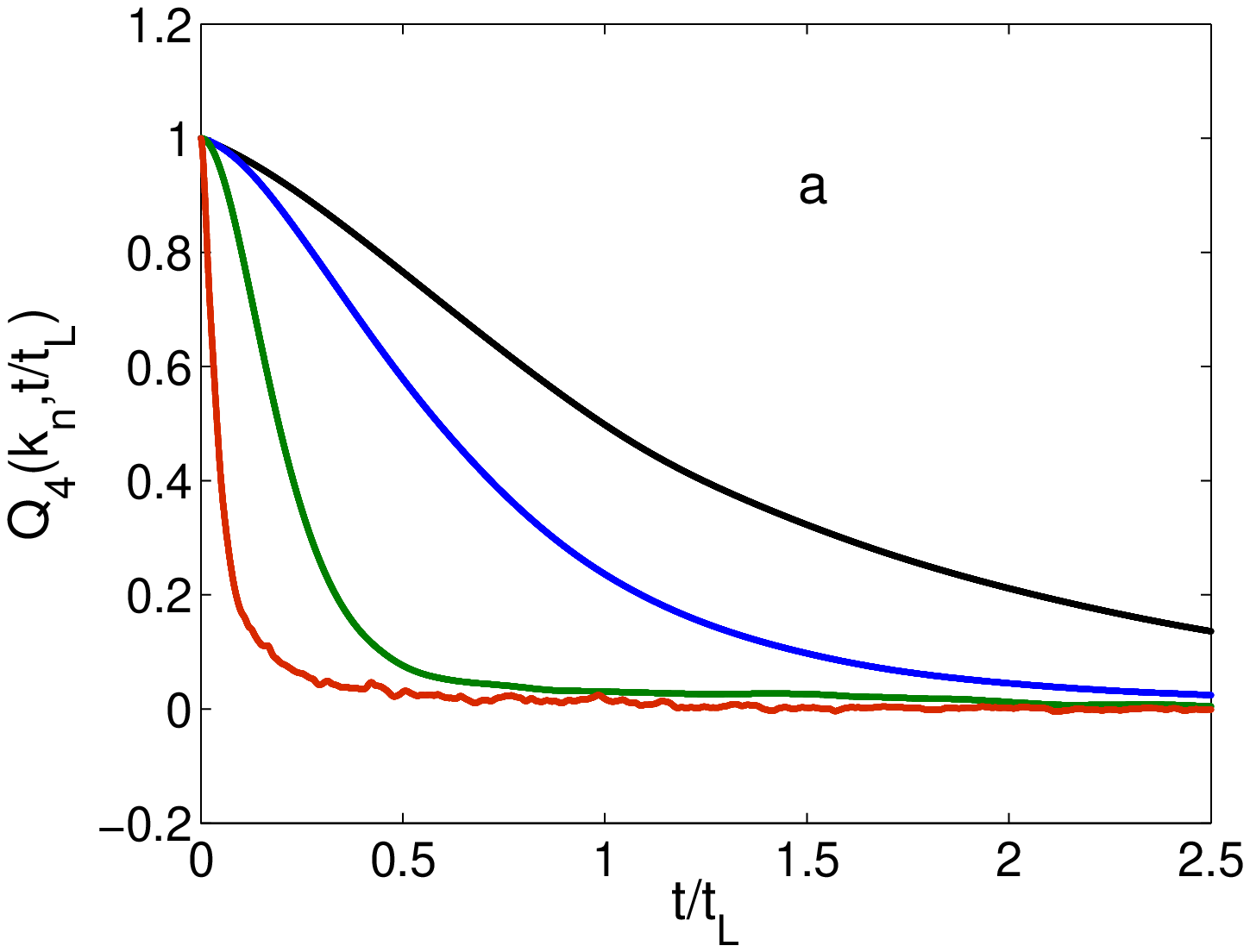}
\includegraphics[height=6cm,width=6cm]{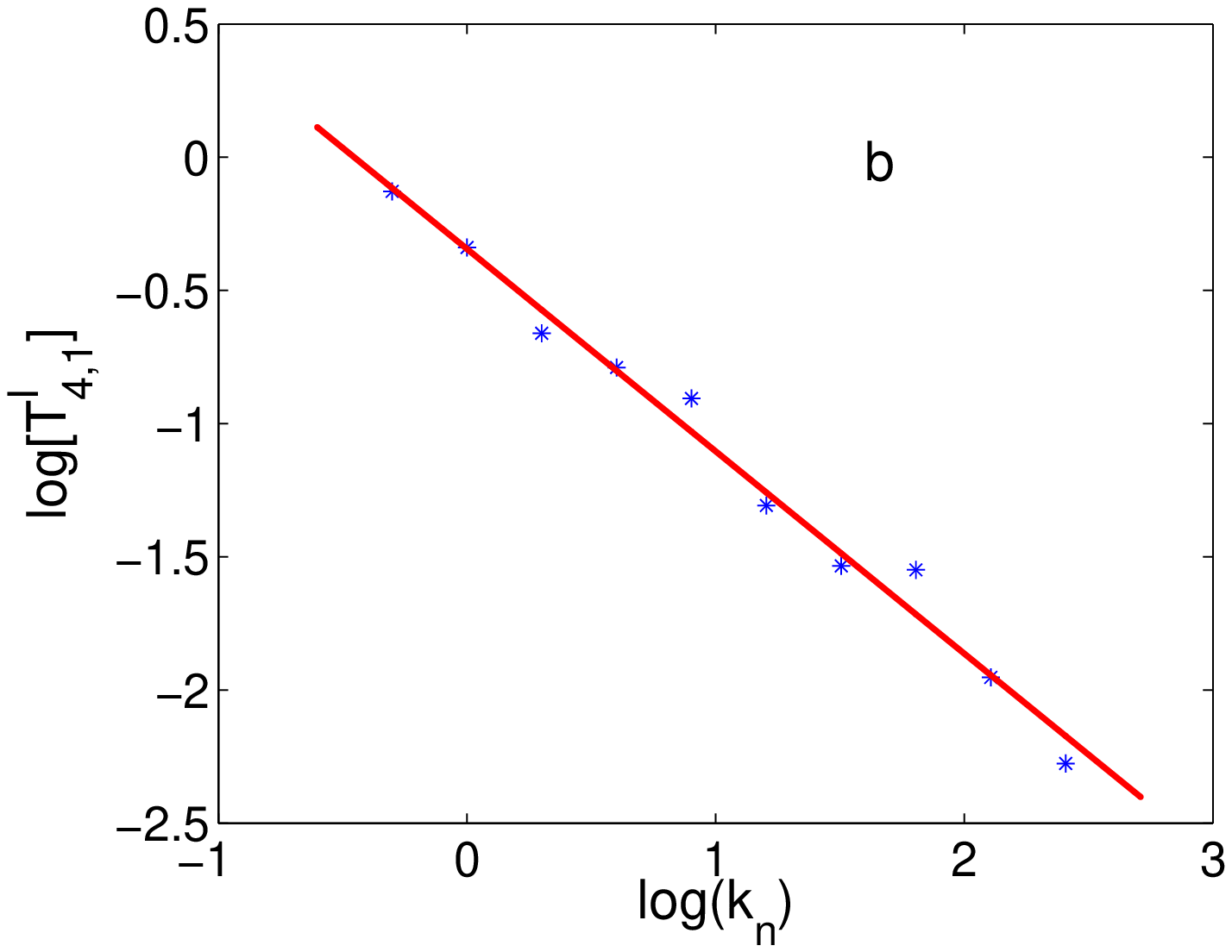}
\caption{(a) Illustrative plots of the normalised, 
fourth-order, time-dependent structure function $Q_4(k_n,t)$ 
versus time $t/t_L$ for shells $n = 3, 4, 7$ and $9$ (from top to 
bottom). (b) A log-log plot of the integral-time scale 
$T^I_{4,1}$ versus $k_n$; the integral-time exponent 
$z_{4,1}^I = 0.728 \pm 0.006$ follows from the slope  
of the line (in the range $4\leq n \leq 12$).}
\label{Q4}
\end{center} 
\end{figure}

To calculate the integral-time scale $T^I_{p,1}$, we evaluate
the integral in equation (\ref{timp}) with $Q_p$ instead of
${\cal F}_p$  and the upper limit replaced by $t_{\mu}$, the time 
at which $Q_p(k_n,t)$ reaches a value $\mu \in [0,1]$. We should, 
of course, use $\mu = 0$, i.e., $t_{\mu} = \infty$, but it is hard 
to obtain well-averaged data for large $t$, so we use
$\mu = 0.6$. We have checked that our results are not affected
significantly if we restrict ourselves to the range $0.4 \leq \mu
\leq 0.7$.  In Figure \ref{Q4}(a) we show representative plots of
$Q_4$ versus time $t/t_L$ for shells $n= 3, 4, 7,$ and $9$
and $\alpha = 2$; from this we obtain $T^I_{4,1}$ as described 
above. The slopes of log-log plots of $T^I_{p,1}$ versus $k_n$ 
now yield $z^I_{p,1}$ as shown in Figure \ref{Q4}(b) for $p=4$. 
The derivative-time scale $T^D_{p,2}$ is calculated by using a 
sixth-order, centred difference scheme. The derivative-time 
exponent $z^D_{p,2}$ is then obtained from slopes of log-log
plots of $T^D_{p,2}$ versus $k_n$.

\begin{figure}
\begin{center}
\includegraphics[height=6cm,width=6cm]{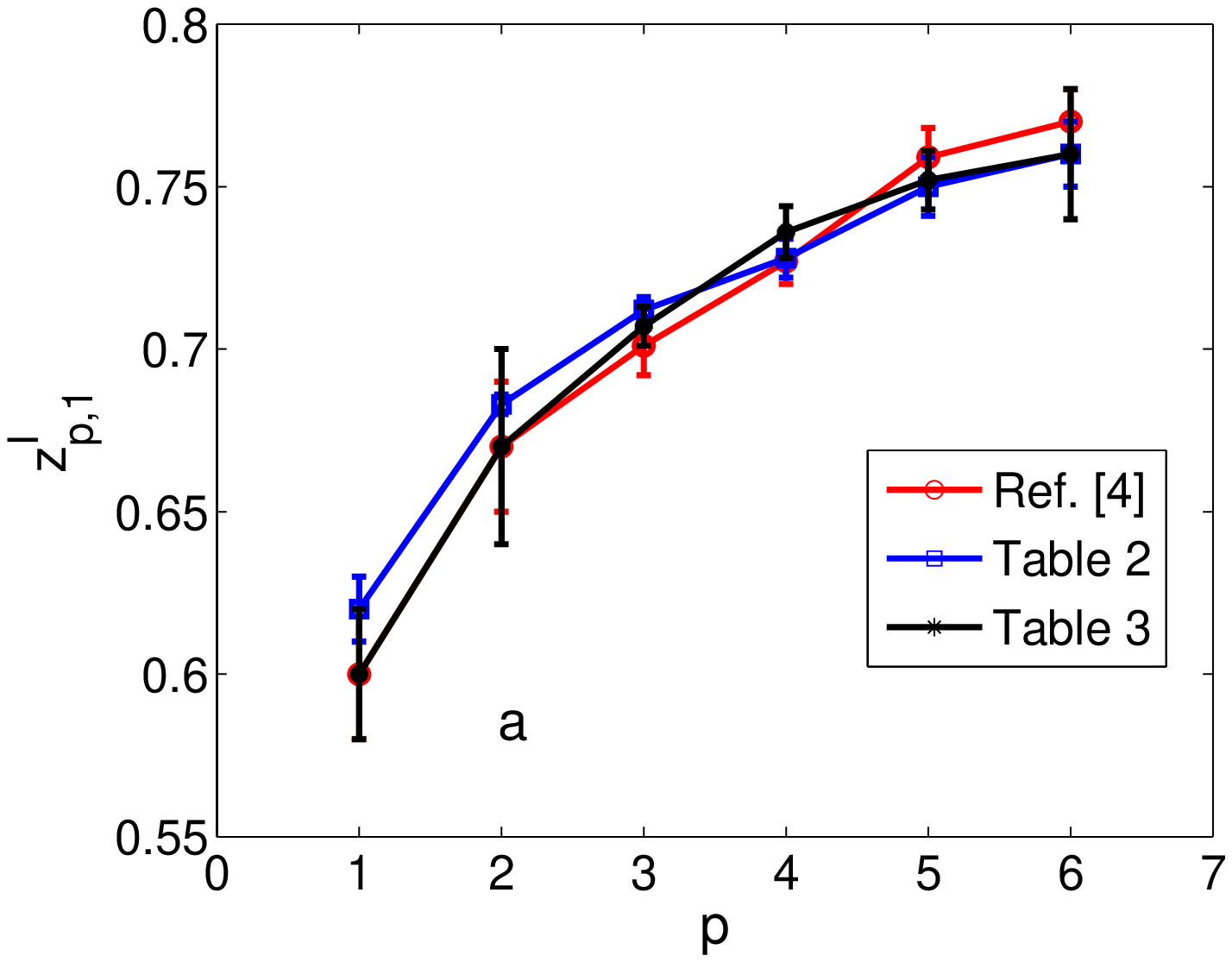}
\includegraphics[height=6cm,width=6cm]{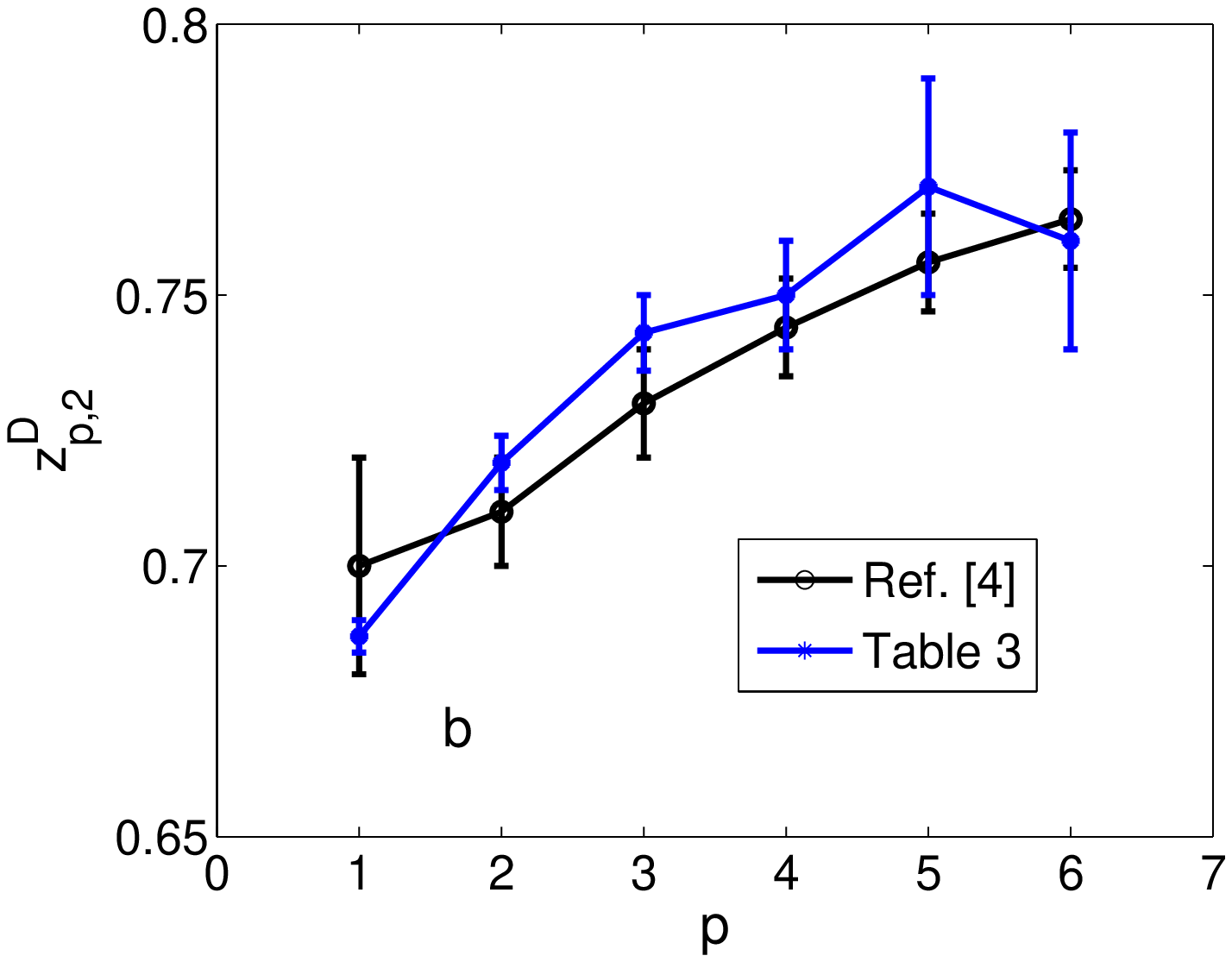}
\caption{Plots of the dynamic multiscaling exponents 
(a) $z^I_{p,1}$ and (b) $z^D_{p,2}$ versus $p$; the error bars 
are obtained as described in the text. These plots 
compare data for statistically steady turbulence in the 
GOY model (from Table 2 in Ref.~\cite{mitra1}) and our data 
for decaying turbulence in the GOY model with normal viscosity 
($\alpha = 0$) and hyperviscosity ($\alpha = 2$); and they
illustrate the universality of dynamic-multiscaling exponents
that we discus in the text.}
\label{comp_integ}
\end{center} 
\end{figure}

Tables 2 and 3 summarise our results for the dynamic-multiscaling
exponents. All these exponents have been calculated for $k_n$
in the inertial range $4\leq n \leq 14$; the exponents $z^D_{p,2}$ 
are shown only for $\alpha = 0$. In Tables 2 and 3
we also give the exponents that we get by using the bridge 
relations (\ref{zipm}) and (\ref{zdpm}) along with the
equal-time exponents given in Table 1; there is good agreement 
between the exponents from our numerical simulations and 
those obtained via the bridge relations.
Moreover, the exponents $z^I_{p,1}$ and $z^D_{p,2}$, 
for $\alpha = 0$, and $z^I_{p,1}$, for $\alpha = 2$, 
which we obtain for decaying turbulence, are equal (within error 
bars) to their counterparts (see Table 2 of Ref.~\cite{mitra1}) 
for statistically steady turbulence. Thus the dynamic-multiscaling 
exponents for the GOY model are universal. Plots of these 
exponents versus $p$ are shown in Figures \ref{comp_integ}(a) and 
\ref{comp_integ}(b). 

\begin{table}
\caption{Column 2: The dynamic multiscaling exponents $z^I_{p,1}$ 
from the bridge relation (\ref{zipm}) and our numerical 
results for $\zeta_p$ (Table 1). Column 3: The same exponents as
in Column 2 but now obtained from our numerical simulations of 
decaying turbulence in the GOY shell model (for the case 
$\alpha = 2$).  Note the agreement between the exponents in 
Columns 2 and 3.}
\label{tab:2}     
\begin{center}
\begin{tabular}{lll}
\hline\noalign{\smallskip}
order-$p$ & $z^I_{p,1}$[Eq.(\ref{zipm})] & $z^I_{p,1}$  \\
\noalign{\smallskip}\hline\noalign{\smallskip}
1 & 0.630 $\pm$ 0.001 & 0.62 $\pm$ 0.01 \\
2 & 0.675 $\pm$ 0.009 & 0.683 $\pm$ 0.003 \\
3 & 0.69 $\pm$ 0.01 & 0.712 $\pm$ 0.004 \\
4 & 0.71 $\pm$ 0.02 & 0.728 $\pm$ 0.006 \\
5 & 0.74 $\pm$ 0.05 & 0.750 $\pm$ 0.009 \\
6 & 0.76 $\pm$ 0.08 & 0.76 $\pm$ 0.01 \\
\noalign{\smallskip}\hline
\end{tabular}
\end{center}
\end{table}

\begin{table}
\caption{Dynamic multiscaling exponents from our simulations of 
decaying turbulence in the GOY model with $\alpha = 0$: The bridge 
relations (\ref{zipm}-\ref{zdpm}) are used along with the 
equal-time exponents $\zeta_p$ for $\alpha = 0$ (Table 1) 
to calculate $z^{I}_{p,1}$ (Column 2) and $z^{D}_{p,2}$ 
(Column 4). The corresponding dynamic-multiscaling exponents
that we obtain from our numerical study of the normalised,
time-dependent structure functions $Q_p$ are given in 
Columns 3 and 5.}   
\label{tab3}
\begin{center}
\begin{tabular}{lllll}
\hline\noalign{\smallskip}
order-$p$ & $z^{I}_{p,1}$[Eq.(\ref{zipm})] & 
$z^{I}_{p,1}$ & $z^{D}_{p,2}[Eq.(\ref{zdpm})]$ & $z^{D}_{p,2}$  \\
\noalign{\smallskip}\hline\noalign{\smallskip}
 1 &  0.620 $\pm$ 0.001 & 0.60 $\pm$ 0.02
    &0.690  $\pm$ 0.006 & 0.687 $\pm$ 0.003
   \\
 2 &  0.671 $\pm$ 0.004 & 0.67 $\pm$ 0.03
   & 0.72 $\pm$ 0.01  & 0.719 $\pm$ 0.005
 \\
 3 &  0.709 $\pm$ 0.008 & 0.707 $\pm$ 0.006
   & 0.74 $\pm$ 0.02  & 0.743 $\pm$ 0.007
  \\
 4 &  0.73 $\pm$ 0.01 & 0.736 $\pm$ 0.008
 & 0.76 $\pm$ 0.03  & 0.75 $\pm$ 0.01   \\

 5 &  0.75 $\pm$ 0.02  & 0.752 $\pm$ 0.009
   &0.77  $\pm$ 0.03  & 0.77 $\pm$ 0.02
  \\
 6 &  0.77 $\pm$ 0.03   & 0.76 $\pm$ 0.02
   & 0.77  $\pm$ 0.03   & 0.76 $\pm$ 0.02
   \\
\noalign{\smallskip}\hline
\end{tabular}
\end{center}
\end{table}

The universality of dynamic exponents also goes through for
models of passive-scalar turbulence as we show in 
Refs.~\cite{mitra2,ray}. It turns out that the Kraichnan 
model~\cite{kraich1,falko} shows simple scaling. However, 
a shell-model version of a passive scalar advected by GOY-model
velocities exhibits dynamic multiscaling with dynamic exponents that 
depend on the degree $M$, defined in equations 
(\ref{timp}-\ref{tdp}), but not on the order $p$.

We hope our detailed numerical simulations of dynamic multiscaling 
in the GOY shell model for fluid turbulence will lead to 
experiments designed to measure time-dependent Lagrangian
or quasi-Lagrangian structure functions in fluid turbulence.  
Advances in experimental techniques have made it possible to
get good data for, say, the acceleration of Lagrangian 
particles~\cite{boden}. We believe such techniques can 
be refined to measure the sorts of time-dependent structure 
functions we have discussed here.

We would like to thank Prasad Perlekar and Ganapati Sahoo for 
discussions, UGC and DST (India) for support, and SERC (IISc) for 
computational resources. One of us (RP) is also part of the 
International Collaboration for Turbulence Research (ICTR).

\end{document}